\documentclass[11pt, letterpaper]{article}
\usepackage{amsmath, amssymb, amsthm, amsfonts, bbm, commath}
\usepackage[inline]{enumitem}
\usepackage{tikz}
\usetikzlibrary{arrows.meta,shapes}
\usepackage{subcaption}
\usepackage{authblk}
\usepackage{graphicx}
\usepackage{hyperref}
\usepackage{xcolor}
\usepackage[ruled]{algorithm2e}
\usepackage{cancel}
\usepackage{multirow}
\usepackage{calc}
\usepackage[letterpaper, margin=1in]{geometry}

\title{
	Asynchronous Byzantine Consensus on Undirected Graphs under Local Broadcast Model
	\thanks{
		This research is supported in part by the National Science Foundation award 1733872 and Toyota InfoTechnology Center. Any opinions, findings, and conclusions or recommendations expressed here are those of the authors and do not necessarily reflect the views of the funding agencies or the U.S. government.
	}
}

\author[1]{
	Muhammad Samir Khan
}

\author[2]{
	Nitin H. Vaidya
}

\affil[1]{
	Department of Computer Science \protect\linebreak
	University of Illinois at Urbana-Champaign \protect\linebreak
	\texttt{mskhan6@illinois.edu} \protect\linebreak
}

\affil[2]{
	Department of Computer Science \protect\linebreak
	Georgetown University \protect\linebreak
	\texttt{nv198@georgetown.edu}
}

\newtheorem{theorem}{Theorem}[section]

\newcommand{\COMMENTS}{}

\ifdefined\COMMENTS
   \newcommand{\comment}[1]{{\color{red}#1}}
\else
   \newcommand{\comment}[1]{}
\fi

\newenvironment{proof_of}[1]{\noindent {\bf Proof of #1:}
	\hspace*{1mm}}{\hspace*{\fill} $\Box$ }

\newcommand{\adjacent}[1]{\overset{}{\rightarrow}}
\newcommand{\notadjacent}[1]{\overset{}{\not \rightarrow}}

\begin{document}
	\maketitle
	
	\section{Introduction}
		In this work we look at Byzantine consensus in asynchronous systems under the \emph{local broadcast} model.
		In the \emph{local broadcast} model \cite{Bhandari:2005:RBR:1073814.1073841, Koo:2004:BRN:1011767.1011807}, a message sent by any node is received identically by \emph{all of its neighbors} in the communication network, preventing a faulty node from transmitting conflicting information to different neighbors.
		Our recent work \cite{Undirected_PODC2019} has shown that in the \emph{synchronous} setting, network connectivity requirements for Byzantine consensus are lower under the local broadcast model as compared to the classical point-to-point communication model.
		Here we show that the same is not true in the asynchronous setting, and the network requirements for Byzantine consensus stays the same under local broadcast as under point-to-point communication model.
		
		A classical result \cite{fischer1982impossibility} shows that it is impossible to reach exact consensus even with a single crash failure in an asynchronous system.
		However, despite asynchrony, approximate Byzantine consensus among $n$ nodes in the presence of $f$ Byzantine faulty nodes is possible in networks with vertex connectivity at least $2f+1$ and $n \ge 3f + 1$ \cite{Dolev:1986:RAA:5925.5931}.
		Motivated by results in the synchronous setting \cite{Undirected_PODC2019}, one might expect a lower connectivity requirement under the local broadcast model.
		In this work we show that, in fact, the network conditions do not change from the point-to-point communication model.
	
	\section{System Model and Notation}
		We represent the communication network by an undirected graph $G = (V, E)$.
		Each node knows the graph $G$.
		Each node $u$ is represented by a vertex $u \in V$.
		We use the terms \emph{node} and \emph{vertex} interchangeably.
		Two nodes $u$ and $v$ are \emph{neighbors} if and only if $uv \in E$ is an edge of $G$.
		
		Each edge $uv$ represents a FIFO link between two nodes $u$ and $v$.
		When a message $m$ sent by node $u$ is received by node $v$, node $v$ knows that $m$ was sent by node $u$.
		We assume the \emph{local broadcast} model wherein a message sent by a node $u$ is received identically and correctly by each node $v$ such that $uv \in E$ (i.e., by each neighbor of $u$)\footnote{Our results apply even for the stronger model where messages must be received at the same time by all the neighbors.}.
		We assume an asynchronous system where the nodes proceed at varying speeds, in the absence of a global clock, and messages sent by a node are received after an unbounded but finite delay\footnote{Our results apply even for the stronger model where messages are received after a known bounded delay as well as (with slight modifications to the proofs) to the case where message delay is unbounded but nodes have a global clock for synchronization.}.
		
		A \emph{Byzantine} faulty node may exhibit arbitrary behavior.
		There are $n$ nodes in the system of which at most $f$ nodes may be Byzantine faulty, where $0 < f < n$\footnote{The case where $f = 0$ is trivial and the case when $n = f$ is not of interest.}.
		We consider the \emph{$\epsilon$-approximate Byzantine consensus problem} where each of the $n$ nodes starts with a \emph{real valued input}, with known upper and lower bounds $U$ and $L$ such that $L < U$ and $U - L > \epsilon > 0$.
		Each node must output a real value satisfying the following conditions.
		\begin{enumerate}[label=\arabic*)]
			\item \textbf{$\epsilon$-Agreement:}
			For any two non-faulty nodes, their output must be within a fixed constant $\epsilon$.
			\item \textbf{Validity:}
			The output of each non-faulty node must be in the convex hull of the inputs of non-faulty nodes.
			\item \textbf{Termination:}
			All non-faulty nodes must decide on their output in finite time which can depend on $U$, $L$, and $\epsilon$.
		\end{enumerate}
		
		Once a node terminates, it takes no further steps.
	
	\section{Impossibility Results}
		In this section we show two impossibility results.
		
		\begin{theorem} \label{theorem number of nodes}
			If there exists an $\epsilon$-approximate Byzantine consensus algorithm under the local broadcast model on an undirected graph $G$ tolerating at most $f$ Byzantine faulty nodes, then $n \ge 3f+1$.
		\end{theorem}
		
		\begin{theorem} \label{theorem connectivity network}
			If there exists an $\epsilon$-approximate Byzantine consensus algorithm under the local broadcast model on an undirected graph $G$ tolerating at most $f$ Byzantine faulty nodes, then $G$ is $(2f+1)$-connected.
		\end{theorem}
		
		Both the proofs follow the state machine based approach \cite{Attiya:2004:DCF:983102, Dolev:1986:RAA:5925.5931, Fischer1986}.
		
		\begin{proof_of}{Theorem \ref{theorem number of nodes}}
			We assume that $G$ is a complete graph; if consensus can not be achieved on a complete graph consisting of $n$ nodes, then it clearly cannot be achieved on a partially connected graph consisting of $n$ nodes.
			Suppose for the sake of contradiction that $n \le 3f$ and there exists an algorithm $\mathcal{A}$ that solves $\epsilon$-approximate Byzantine consensus in an asynchronous system under the local broadcast model.
			Then there exists a partition $(A, B, C)$ of $V$ such that $\abs{A}, \abs{B}, \abs{C} \le f$.
			Since $n > f \ge 1$, we can ensure that both $A$ and $B$ are non-empty.
			Algorithm $\mathcal{A}$ outlines a procedure $\mathcal{A}_u$ for each node $u$ that describes $u$'s state transitions.
			
			We first create a network $\mathcal{G}$ to model behavior of nodes in $G$ in two different executions $E_1$ and $E_2$, which we will describe later.
			Figure \ref{figure number of nodes} depicts $\mathcal{G}$.
			The network $\mathcal{G}$ consists of two copies of each node in $C$, denoted by $C_{\operatorname{crash}}$ and $C_{\operatorname{slow}}$, and a single copy of each of the remaining nodes.
			For each node $u$ in $G$, we have the following cases to consider:
			\begin{enumerate}[label=\arabic*)]
				\item If $u \in A$, then there is a single copy of $u$ in $\mathcal{G}$.
				With a slight abuse of terminology, we denote the copy by $u$ as well.
				
				\item If $u \in B$, then there is a single copy of $u$ in $\mathcal{G}$.
				With a slight abuse of terminology, we denote the copy by $u$ as well.
				
				\item If $u \in C$, then there are two copies of $u$ in $\mathcal{G}$.
				We denote the two copies by $u_{\operatorname{crash}} \in C_{\operatorname{crash}}$ and $u_{\operatorname{slow}} \in C_{\operatorname{slow}}$.
			\end{enumerate}
			
			For each edge $uv \in E(G)$, we create edges in $\mathcal{G}$ as follows:
			\begin{enumerate}[label=\arabic*)]
				\item If $u, v \in A \cup B$, then there is an edge between the corresponding copy of $u$ and $v$ in $\mathcal{G}$.
				
				\item If $u \not \in C, v \in C$, then there is a single edge $u v_{\operatorname{slow}}$ in $\mathcal{G}$.
				
				\item If $u, v \in C$, then there is an edge $u_{\operatorname{crash}} v_{\operatorname{crash}}$ and an edge $u_{\operatorname{slow}} v_{\operatorname{slow}}$ in $\mathcal{G}$.
			\end{enumerate}
			
			Note that the edges in $G$ and $\mathcal{G}$ are both undirected.
			Observe that the structure of $\mathcal{G}$ ensures the following property.
			For each edge $uv$ in the original graph $G$, each copy of $u$ receives messages from at most one copy of $v$ in $\mathcal{G}$.
			This allows us to create an algorithm for $\mathcal{G}$ corresponding to $\mathcal{A}$ by having each copy $u_i \in \mathcal{G}$ of node $u \in G$ run $\mathcal{A}_u$.
			
			The nodes in $C_{\operatorname{crash}}$ start off in a crashed state and never take any steps.
			The nodes in $C_{\operatorname{slow}}$ are ``slow'' and start taking steps after time $\Delta$, where the value of $\Delta$ will be chosen later.
			
			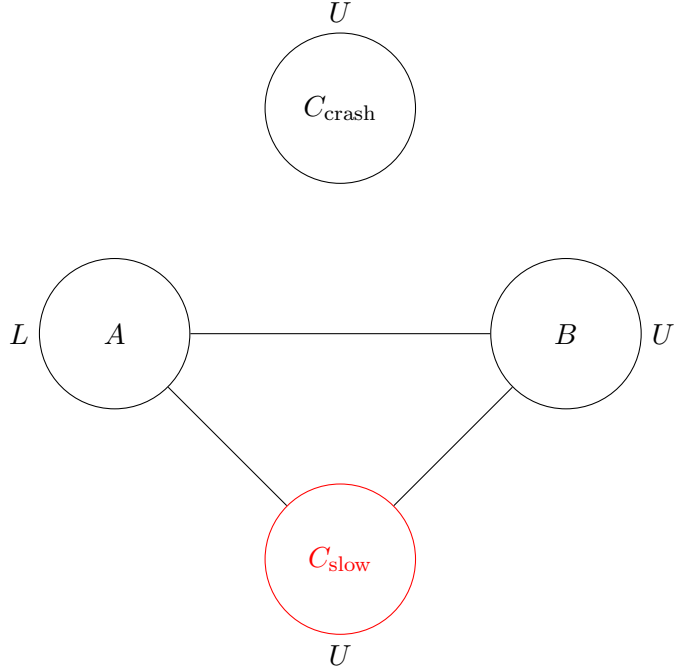
\begin{figure}
				\centering
				\begin{tikzpicture}
				\node[draw, circle, minimum size=2cm, label={left:$L$}] at (0, 0) (A) {$A$};
				\node[draw, circle, minimum size=2cm, label={right:$U$}] at (6, 0) (B) {$B$};
				\node[draw, circle, minimum size=2cm, label={above:$U$}] at (3, 3) (C_crash) {$C_{\operatorname{crash}}$};
				\node[draw, circle, minimum size=2cm, label={below:$U$}, red] at (3, -3) (C_slow) {$C_{\operatorname{slow}}$};
				
				\draw[-] (A) to (B);
				\draw[-] (B) to (C_slow);
				\draw[-] (C_slow) to (A);
				\end{tikzpicture}
				\caption{
					Network $\mathcal{G}$ to model executions $E_1$ and $E_2$ in proof of Theorem \ref{theorem number of nodes}.
					Edges within the sets are not shown while edges between sets are depicted as single edges.
					The labels adjacent to the sets are the corresponding inputs in execution $\mathcal{E}$.
				}
				\label{figure number of nodes}
			\end{figure}
			
			Consider an execution $\mathcal{E}$ of the above algorithm on $\mathcal{G}$ as follows.
			Each node in $A$ has input $L$ and each node in $B \cup C_{\operatorname{slow}} \cup C_{\operatorname{crash}}$ has input $U$.
			Observe that it is not guaranteed that nodes in $\mathcal{G}$ will satisfy any of the conditions of $\epsilon$-approximate Byzantine consensus, including the termination property.
			We will show that the algorithm does indeed terminate but the output of the nodes do not satisfy the validity condition, which will give us the desired contradiction.
			We use $\mathcal{E}$ to describe two executions $E_1$ and $E_2$ of $\mathcal{A}$ on the original graph $G$ as follows.
			\begin{enumerate}[label=$E_{\arabic*}$:,labelindent=0pt]
				\item
				$C$ is the set of faulty nodes which crash immediately at the start of the execution.
				Each node in $A$ has input $L$ while all other nodes have input $U$.
				Since $\mathcal{A}$ solves $\epsilon$-approximate Byzantine consensus on $G$, nodes in $A \cup B$ reach $\epsilon$-agreement and terminate within some finite time, without receiving any messages from nodes in $C$.
				We set $\Delta$ for the delay above for $C_{\operatorname{slow}}$ to be this value.
				Since $U - L > \epsilon$, the outputs of (non-faulty) nodes in $A \cup B$ are either not $U$ or not $L$.
				WLOG we assume that the outputs are not $U$\footnote{For the other case, we can switch the faulty set in $E_2$ to $B$ and change the input of $C_{\operatorname{slow}}$ to be $L$}.
				Note that the behavior of non-faulty nodes in $A$ and $B$ \emph{for the first $\Delta$ time period} is modeled by the corresponding (copies of) nodes in $\mathcal{G}$, while the behavior of the (crashed) faulty nodes is captured by $C_{\operatorname{crash}}$.
				
				\item
				$A$ is the set of Byzantine faulty nodes.
				A faulty node broadcasts the same messages as the corresponding node in $\mathcal{G}$ in execution $\mathcal{E}$.
				Each node in $A$ has input $L$ while all other nodes have input $U$.
				The output of the non-faulty nodes will be described later.
				The behavior of nodes (both faulty and non-faulty) in $A$ and $B$ is modeled by the corresponding (copies of) nodes in $\mathcal{G}$, while the behavior of the (non-faulty) nodes in $C$ is captured by $C_{\operatorname{slow}}$.
			\end{enumerate}
			
			Due to the behavior of nodes in $A$ and $B$ in $E_1$, each of the corresponding copies in $\mathcal{G}$ decides on a value distinct from $U$ and terminates within time $\Delta$ in execution $\mathcal{E}$.
			Therefore, the behavior of nodes in $A$ and $B$ is completely captured by the corresponding copies in $\mathcal{E}$.
			It follows that in $E_2$, nodes in $B$ have outputs other than $U$.
			However, all non-faulty nodes have input $U$ in $E_2$.
			Recall that, by construction, $B$ is non-empty.
			This violates validity, a contradiction.
		\end{proof_of}\\
		
		\begin{proof_of}{Theorem \ref{theorem connectivity network}}
			Suppose for the sake of contradiction that $G$ is not $(2f+1)$-connected and there exists an algorithm $\mathcal{A}$ that solves $\epsilon$-approximate Byzantine consensus in an asynchronous system under the local broadcast model on $G$.
			Then there exists a vertex cut $C$ of $G$ of size at most $2f$ with a partition $(A, B, C)$ of $V$ such that $A$ and $B$ (both non-empty) are disconnected in $G - C$ (so there is no edge between a node in $A$ and a node in $B$).
			Since $\abs{C} \le 2f$, there exists a partition $(C^1, C^2)$ of $C$ such that $\abs{C^1}, \abs{C^2} \le {f}$.
			Algorithm $\mathcal{A}$ outlines a procedure $\mathcal{A}_u$ for each node $u$ that describes $u$'s state transitions.
			
			We first create a network $\mathcal{G}$ to model behavior of nodes in $G$ in three different executions $E_1$, $E_2$, and $E_3$, which we will describe later.
			Figure \ref{figure connectivity network} depicts $\mathcal{G}$.
			The network $\mathcal{G}$ consists of three copies of each node in $C^1$, two copies of each node in $A$ and $B$, and a single copy of each node in $C^2$.
			We denote the three sets of copies of $C_1$ by $C^1_{\operatorname{crash}}$, $C^1_L$, and $C^1_U$.
			We denote the two sets of copies of $A$ (resp. $B$) by $A_L$ and $A_U$ (resp. $B_L$ and $B_U$).
			For each edge $uv \in E(G)$, we create edges in $\mathcal{G}$ as follows:
			\begin{enumerate}[label=\arabic*)]
				\item If $u, v \in A$ (resp. $\in B$), then there are two copies of $u$ and $v$, $u_L, v_L \in A_L$ (resp. $\in B_L$) and $u_U, v_U \in A_U$ (resp. $\in B_U$).
				There is an edge $u_L v_L$ and an edge $u_U v_U$ in $\mathcal{G}$.
				
				\item If $u, v \in C^1$, then there are three copies $u_L, v_L \in C^1_L$, $u_U, v_U \in C^1_U$, and $u_{\operatorname{crash}}, v_{\operatorname{crash}} \in C^1_{\operatorname{crash}}$ of $u$ and $v$.
				There are edges $u_L v_L$, $u_U v_U$, $u_{\operatorname{crash}} v_{\operatorname{crash}}$ in $\mathcal{G}$.
				
				\item If $u, v \in C^2$, then there is an edge $uv$ between the corresponding copies in $\mathcal{G}$.
				
				\item If $u \in C^1, v \in C^2$, then there are three copies $u_L \in C^1_L$, $u_U  \in C^1_U$, and $u_{\operatorname{crash}}  \in C^1_{\operatorname{crash}}$ of $u$, and a single copy of $v$.
				There is an undirected edge $u_U v$ and a directed edge $\overrightarrow{v u_L}$ in $\mathcal{G}$.
				
				\item If $u \in A, v \in C^1$, then there are two copies $u_L \in A_L$ and $u_U \in A_U$ of $u$, and three copies $v_L \in C^1_L$, $v_U  \in C^1_U$, and $v_{\operatorname{crash}}  \in C^1_{\operatorname{crash}}$ of $v$.
				There are two undirected edges $u_L v_L$ and $u_U v_U$ in $\mathcal{G}$.
				
				\item If $u \in B, v \in C^1$, then there are two copies $u_L \in B_L$ and $u_U \in B_U$ of $u$, and three copies $v_L \in C^1_L$, $v_U  \in C^1_U$, and $v_{\operatorname{crash}}  \in C^1_{\operatorname{crash}}$ of $v$.
				There are two undirected edges $u_L v_L$ and $u_U v_U$ in $\mathcal{G}$.
				
				\item If $u \in A, v \in C^2$, then there are two copies $u_L \in A_L$ and $u_U \in A_U$ of $u$, and a single copy of $v$.
				There is an undirected edge $u_L v$ and a directed edge $\overrightarrow{v u_U}$ in $\mathcal{G}$.
				
				\item If $u \in B, v \in C^2$, then there are two copies $u_L \in B_L$ and $u_U \in B_U$ of $u$, and a single copy of $v$.
				There is an undirected edge $u_U v$ and a directed edge $\overrightarrow{v u_L}$ in $\mathcal{G}$.
			\end{enumerate}
			
			$\mathcal{G}$ has some directed edges.
			We describe their behavior next.
			We denote a directed edge from $u$ to $v$ as $\overrightarrow{u v}$.
			All message transmissions in $\mathcal{G}$ are via local broadcast, as follows.
			When a node $u$ in $\mathcal{G}$ transmits a message, the following nodes receive this message identically: each node with whom $u$ has an undirected edge and each node to whom there is an edge directed away from $u$.
			Note that a directed edge $e = \overrightarrow{u v}$ behaves differently for $u$ and $v$.
			All messages sent by $u$ are received by $v$.
			No message sent by $v$ is received by $u$.
			Observe that with this behavior of directed edges, the structure of $\mathcal{G}$ ensures the following property.
			For each edge $uv$ in the original graph $G$, each copy of $u$ receives messages from at most one copy of $v$ in $\mathcal{G}$.
			This allows us to create an algorithm for $\mathcal{G}$ corresponding to $\mathcal{A}$ by having each copy $u_i \in \mathcal{G}$ of node $u \in G$ run $\mathcal{A}_u$.
			
			The nodes in $C^1_{\operatorname{crash}}$ start off in a crashed state and never take any steps.
			The nodes in $C^1_L$ and $C^1_U$ are ``slow'' and start taking steps after time $\Delta$, where the value of $\Delta$ will be chosen later.
			
			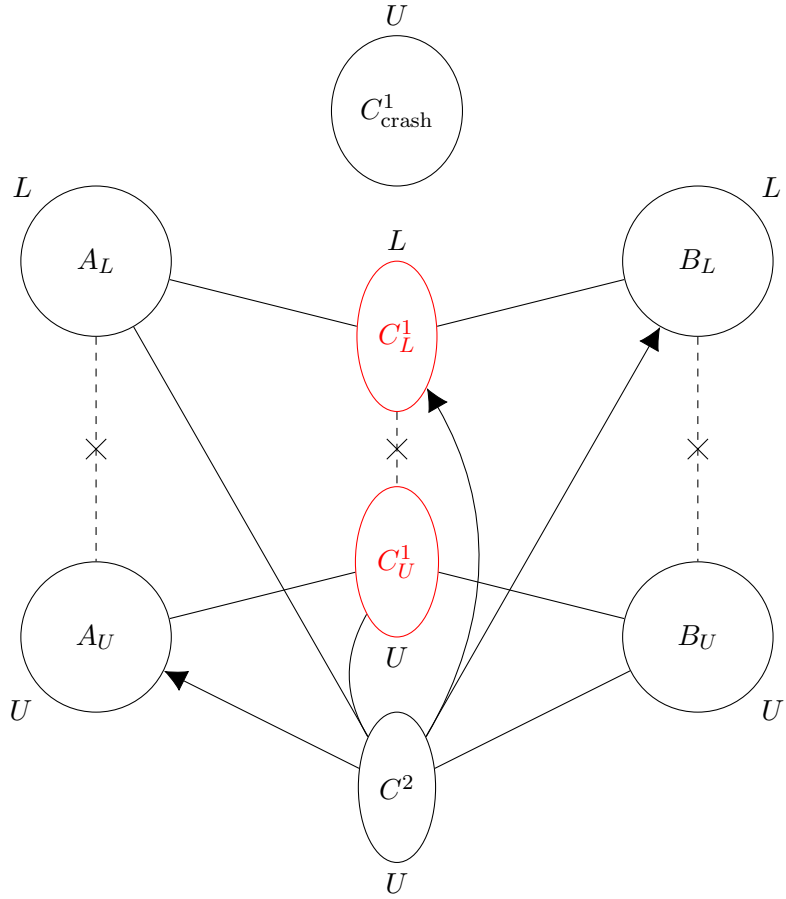
\begin{figure}
				\centering
				\begin{tikzpicture}
				\node[draw, circle, minimum size=2cm, label={above left:$L$}] at (0, 2) (A_0) {$A_L$};
				\node[draw, circle, minimum size=2cm, label={above right:$L$}] at (8, 2) (B_0) {$B_L$};
				\node[draw, circle, minimum size=2cm, label={below left:$U$}] at (0, -3) (A_1) {$A_U$};
				\node[draw, circle, minimum size=2cm, label={below right:$U$}] at (8, -3) (B_1) {$B_U$};
				\node[draw, ellipse, minimum height=2cm, label={above:$U$}] at (4, 4) (C^1_c) {$C^1_{\operatorname{crash}}$};
				\node[draw, ellipse, minimum height=2cm, label={above:$L$}, red] at (4, 1) (C^1_0) {$C^1_L$};
				\node[draw, ellipse, minimum height=2cm, label={below:$U$}, red] at (4, -2) (C^1_1) {$C^1_U$};
				\node[draw, ellipse, minimum height=2cm, label={below:$U$}] at (4, -5) (C^2) {$C^2$};
				
				\draw[dashed] (A_0) to node[cross out,draw,solid]{} (A_1);
				\draw[dashed] (B_0) to node[cross out,draw,solid]{} (B_1);
				\draw[dashed] (C^1_0) to node[cross out,draw,solid]{} (C^1_1);
				
				\draw[-{Latex[width=3mm,length=3mm]}] (C^2) to (B_0);
				\draw[-{Latex[width=3mm,length=3mm]}] (C^2) to (A_1);
				\draw[-{Latex[width=3mm,length=3mm]},bend right] (C^2) to (C^1_0);
				
				\draw[-] (C^1_0) to (A_0);
				\draw[-] (C^1_0) to (B_0);
				\draw[-] (C^2) to (A_0);
				\draw[-] (C^2) to (B_1);
				\draw[-] (C^1_1) to (A_1);
				\draw[-] (C^1_1) to (B_1);
				
				\draw[-,bend right] (C^1_1) to (C^2);
				\end{tikzpicture}
				\caption{
					Network $\mathcal{G}$ to model executions $E_1$, $E_2$, and $E_3$ in proof of Theorem \ref{theorem connectivity network}.
					Edges within the sets are not shown while edges between sets are depicted as single edges.
					The crossed dotted lines emphasize that there are no edges between the corresponding sets.
					The labels adjacent to the sets are the corresponding inputs in execution $\mathcal{E}$.
				}
				\label{figure connectivity network}
			\end{figure}
			
			Consider an execution $\mathcal{E}$ of the above algorithm on $\mathcal{G}$ as follows.
			Each node in $A_L \cup B_L \cup C^1_L$ has input $L$ and all other nodes have input $U$.
			Observe that it is not guaranteed that nodes in $\mathcal{G}$ will satisfy any of the conditions of $\epsilon$-approximate Byzantine consensus, including the termination property.
			We will show that the algorithm does indeed terminate but nodes do not reach $\epsilon$-agreement in $\mathcal{G}$, which will be useful in deriving the desired contradiction.
			We use $\mathcal{E}$ to describe three executions $E_1$, $E_2$, and $E_3$ of $\mathcal{A}$ on the original graph $G$ as follows.
			\begin{enumerate}[label=$E_{\arabic*}$:,labelindent=0pt]
				\item
				$C^1$ is the set of faulty nodes which crash immediately at the start of the execution.
				Each node in $A$ has input $L$ while all other nodes have input $U$.
				Since $\mathcal{A}$ solves $\epsilon$-approximate Byzantine consensus on $G$, nodes in $A \cup B \cup C^2$ reach $\epsilon$-agreement and terminate within some finite time, without receiving any messages from nodes in $C^1$.
				We set $\Delta$ for the delay above for $C^1_L$ and $C^1_U$ to be this value.
				The output of the non-faulty nodes will be described later.
				Note that the behavior of non-faulty nodes in $A$, $B$, and $C^2$ \emph{for the first $\Delta$ time period} is modeled by the corresponding (copies of) nodes in $A_L$, $B_U$, and $C^2$ respectively, while the behavior of the (crashed) faulty nodes is captured by $C^1_{\operatorname{crash}}$.
				
				\item
				$C^2$ is the set of faulty nodes.
				A faulty node broadcasts the same messages as the corresponding node in $\mathcal{G}$ in execution $\mathcal{E}$.
				All non-faulty nodes have input $L$.
				The behavior of non-faulty nodes in $A$, $B$, $C^1$ is modeled by the corresponding (copies of) nodes in $A_L$, $B_L$, and $C^1_L$ respectively, while the behavior of the faulty nodes is captured by $C^2$.
				Since $\mathcal{A}$ solves $\epsilon$-approximate Byzantine consensus on $G$, nodes in $A \cup B \cup C^1$ decide on output $L$.
				
				\item
				$C^2$ is the set of faulty nodes.
				A faulty node broadcasts the same messages as the corresponding node in $\mathcal{G}$ in execution $\mathcal{E}$.
				All non-faulty nodes have input $U$.
				The behavior of non-faulty nodes in $A$, $B$, $C^1$ is modeled by the corresponding (copies of) nodes in $A_U$, $B_U$, and $C^1_U$ respectively, while the behavior of the faulty nodes is captured by $C^2$.
				Since $\mathcal{A}$ solves $\epsilon$-approximate Byzantine consensus on $G$, nodes in $A \cup B \cup C^1$ decide on output $U$.
			\end{enumerate}
			
			Due to the output of nodes in $A$ and $B$ in $E_1$, the nodes in $A_L$ and $B_U$ decide on an output within time $\Delta$ in execution $\mathcal{E}$.
			Therefore, the behavior of nodes in $A$ and $B$ in $E_1$ is completely captured by the corresponding nodes in $A_L$ and $B_U$ in $\mathcal{E}$.
			Now, due to the output of nodes in $A$ in $E_2$, the nodes in $A_L$ output $L$ in $\mathcal{E}$.
			Similarly, due to the output of nodes in $B$ in $E_3$, the nodes in $B_U$ output $U$ in $\mathcal{E}$.
			It follows that in $E_1$, nodes in $A$ have output $L$ while nodes in $B$ have output $U$.
			Recall that, by construction, both $A$ and $B$ are non-empty.
			This violates $\epsilon$-agreement, a contradiction.
		\end{proof_of}
	
	\section{Summary}
		In \cite{Undirected_PODC2019} we showed that network requirements are lower for Byzantine consensus in synchronous systems under the local broadcast model, as compared with the point-to-point communication model.
		One might expect a lower connectivity requirement in the asynchronous setting as well.
		In this work, we have presented two impossibility results in Theorems \ref{theorem number of nodes} and \ref{theorem connectivity network} that show that local broadcast does not help improve the network requirements in asynchronous systems.
	
	\nocite{Khan2019ExactBC}
	\bibliographystyle{abbrv}
	\bibliography{bib}

\begin{thebibliography}{1}

\bibitem{Attiya:2004:DCF:983102}
H.~Attiya and J.~Welch.
\newblock {\em Distributed Computing: Fundamentals, Simulations and Advanced
  Topics}.
\newblock John Wiley \& Sons, Inc., USA, 2004.

\bibitem{Bhandari:2005:RBR:1073814.1073841}
V.~Bhandari and N.~H. Vaidya.
\newblock On reliable broadcast in a radio network.
\newblock In {\em Proceedings of the Twenty-fourth Annual ACM Symposium on
  Principles of Distributed Computing}, PODC '05, pages 138--147, New York, NY,
  USA, 2005. ACM.

\bibitem{Dolev:1986:RAA:5925.5931}
D.~Dolev, N.~A. Lynch, S.~S. Pinter, E.~W. Stark, and W.~E. Weihl.
\newblock Reaching approximate agreement in the presence of faults.
\newblock {\em J. ACM}, 33(3):499--516, May 1986.

\bibitem{Fischer1986}
M.~J. Fischer, N.~A. Lynch, and M.~Merritt.
\newblock Easy impossibility proofs for distributed consensus problems.
\newblock {\em Distributed Computing}, 1(1):26--39, Mar 1986.

\bibitem{fischer1982impossibility}
M.~J. Fischer, N.~A. Lynch, and M.~S. Paterson.
\newblock Impossibility of distributed consensus with one faulty process.
\newblock Technical report, Massachusetts Inst of Tech Cambridge lab for
  Computer Science, 1982.

\bibitem{Undirected_PODC2019}
M.~S. Khan, S.~S. Naqvi, and N.~H. Vaidya.
\newblock Exact byzantine consensus on undirected graphs under local broadcast
  model.
\newblock In {\em Proceedings of the 2019 ACM Symposium on Principles of
  Distributed Computing}, PODC '19, pages 327--336, New York, NY, USA, 2019.
  ACM.

\bibitem{Khan2019ExactBC}
M.~S. Khan, S.~S. Naqvi, and N.~H. Vaidya.
\newblock Exact byzantine consensus on undirected graphs under local broadcast
  model.
\newblock {\em CoRR}, abs/1903.11677, 2019.

\bibitem{ArxivTightReport}
M.~S. Khan and N.~H. Vaidya.
\newblock Byzantine consensus under local broadcast model: Tight sufficient
  condition.
\newblock {\em CoRR}, abs/1901.03804, 2019.

\bibitem{Koo:2004:BRN:1011767.1011807}
C.-Y. Koo.
\newblock Broadcast in radio networks tolerating byzantine adversarial
  behavior.
\newblock In {\em Proceedings of the Twenty-third Annual ACM Symposium on
  Principles of Distributed Computing}, PODC '04, pages 275--282, New York, NY,
  USA, 2004. ACM.

\end{thebibliography}
	
	\appendix
\end{document}